# Ferro-ionic States and Domains Morphology in $Hf_xZr_{1-x}O_2$ Nanoparticles

Eugene A. Eliseev[1], Sergei V. Kalinin[2*], Anna N. Morozovska[3†],

[1] Frantsevich Institute for Problems in Materials Science, National Academy of Sciences of Ukraine, 3, str. Omeliana Pritsaka, 03142 Kyiv, Ukraine

[2] Department of Materials Science and Engineering, University of Tennessee, Knoxville, TN, 37996, USA

[3] Institute of Physics, National Academy of Sciences of Ukraine, 46, pr. Nauky, 03028 Kyiv, Ukraine

Unique polar properties of nanoscale hafnia-zirconia oxides ($Hf_xZr_{1-x}O_2$) are of great interest for condensed matter physics, nanophysics and advanced applications. These properties are connected (at least partially) to the ionic-electronic and electrochemical phenomena at the hafnia surface, interfaces and/or internal grain boundaries. Here we calculated the phase diagrams, dielectric permittivity, spontaneous polar and antipolar ordering, and domain structure morphology in $Hf_xZr_{1-x}O_2$ nanoparticles covered by ionic-electronic charge, originated from the surface electrochemical adsorption. We revealed that the ferro-ionic coupling supports the polar long-range order in the nanoscale $Hf_xZr_{1-x}O_2$, induces and/or enlarges the stability region of the labyrinthine domains towards smaller sizes and smaller environmental dielectric constant at low concentrations of the surface ions, and causes the transition to the single-domain ferro-ionic state at high concentrations of the surface ions. We predict that the labyrinthine domain states, being multiple-degenerated, may significantly affect the emergence of the negative differential capacitance state in the nanograined/nanocrystalline $Hf_xZr_{1-x}O_2$ films.

---

[*] corresponding author, e-mail: sergei2@utk.edu

[†] corresponding author, e-mail: anna.n.morozovska@gmail.com

# I. INTRODUCTION

Unique polar properties, such as strong and reversible ferroelectricity and/or antiferroelectricity, of nanoscale fluorite and wurtzite ferroelectrics are of great fundamental interest for condensed matter physics, nanophysics and promising for the next generation of ferroelectric memory and nano-electronic devices [1, 2]. These properties, which are absent in corresponding bulk materials, are connected (at least partially) to the ionic-electronic and electrochemical phenomena [3, 4] at their surfaces, interfaces and/or internal grain boundaries [5, 6]. Notably that the electrochemical effects and ionic density of states at the internal interfaces should offer very similar physical effects.

Si-compatible and lead-free ferroelectric-antiferroelectric hafnium-zirconium oxide ($Hf_xZr_{1-x}O_2$) thin films are suitable candidates for integration with modern silicon-based FeRAMs and FETs [7, 8]. The unusual ferroelectric and/or antiferroelectric properties observed in $Hf_xZr_{1-x}O_2$ films of thickness 5 – 30 nm and Hf content $0.3 \leq x \leq 0.7$ [5, 6] originate from the polar orthorhombic phase $Pca2_1$ (and maybe from the $Pmn2_1$ phase with a very close energy [8]), which is unstable in the bulk monoclinic hafnium-zirconium oxides. Polar properties of $Hf_xZr_{1-x}O_2$ thin films are critically sensitive the size effects [9], as well as to the electrode material, elastic strains, deposition and/or annealing conditions [10] and doping [11, 12]. Not less important can be role of surface and grain boundary energies [5], oxygen vacancies [13, 14], surface and bulk electrochemical states [3, 4].

In comparison with thin epitaxial films, the role of the Zr doping, size, screening and surface effects in $Hf_xZr_{1-x}O_2$ polycrystalline films and nanoparticles are poorly studied (both theoretically and experimentally). There are only several papers, which demonstrate the possibilities of controllable synthesis of $Hf_xZr_{1-x}O_{2-y}$ nanoparticles, which are mostly monoclinic or tetragonal [15, 16, 17, 18], and rarely orthorhombic [19, 20]. The polycrystalline $Hf_xZr_{1-x}O_2$ films can be imagined as an ensemble of single-crystalline nanoparticles separated by grain boundaries, which capture screening charges, charged vacancies and/or ions, thus having finite densities of electronic and ionic states.

In the case of the $Hf_xZr_{1-x}O_2$ nanoparticles (or nanograins) in different environments the Stephenson-Highland (SH) ionic adsorption can occur at their surface [21, 22]. Within the SH model the dependence of the surface charge density on electric potential excess is controlled by the concentration of positive and negative charges at the surface of the nanoparticle. The synergy of SH approach with the Landau-Ginzburg-Devonshire-Kittel (LGDK) phenomenology for the description of coupled polar and antipolar long-range orders in ferroics predicted theoretically [23] and explained experimentally observed [4, 24] versatile ferro-ionic and anti-ferro-ionic states in ferroelectric [25, 26, 27] and antiferroelectric [28, 29] thin films, ferroelectric and ferrielectric nanoparticles [30, 31].

Using a combination of the "effective" LGDK approach proposed for the nanosized $Hf_xZr_{1-x}O_{2-y}$ [32, 20] and the SH approach, in this work we calculate the phase diagrams, dielectric permittivity, spontaneous polar and antipolar orderings, and domain structure morphology in spherical $Hf_xZr_{1-x}O_2$ nanoparticles, which surface interacts with electrochemically active environment.

## II. THEORETICAL MODELLING
### A. Problem formulation

To determine the phase diagrams of the stress-free $Hf_xZr_{1-x}O_2$ nanoparticles we use the LGDK-SH approach. Corresponding free energy functional $F$ includes a bulk part, $F_{bulk}$, a polarization gradient energy contribution, $F_{grad}$, an electrostatic contribution, $F_{el}$, and a surface energy, $F_S$:

$$F = F_{bulk} + F_{grad} + F_{el} + F_S, \qquad (1)$$

where the energy contributions are

$$F_{bulk} = \int_V dV \left( \frac{1}{2} a_{Pi} P_i^2 + \frac{1}{2} a_{Ai} A_i^2 + \frac{1}{4} \beta_{ij} \left( P_i^2 P_j^2 + A_i^2 A_j^2 \right) + \frac{1}{2} \eta_{ij} P_i^2 A_j^2 \right), \qquad (2a)$$

$$F_{grad} = \int_V dV \frac{g_{ijkl}}{2} \left( \frac{\partial P_i}{\partial x_j} \frac{\partial P_k}{\partial x_l} + \frac{\partial A_i}{\partial x_j} \frac{\partial A_k}{\partial x_l} \right), \qquad (2b)$$

$$F_{el} = - \int_V dV \left( P_i E_i + \frac{\varepsilon_0 \varepsilon_b}{2} E^2 \right), \qquad (2c)$$

$$F_S = \frac{1}{2} \int_S dS \; (c_P P^2 + c_A A^2). \qquad (2d)$$

Here $V$ is the volume and $S$ is the surface of the $Hf_xZr_{1-x}O_2$ nanoparticle. The Einstein summation rule over the repeated Cartesian subscripts is used. Polar and antipolar order parameters are $\vec{P} = (P_1, P_2, P_3)$ and $\vec{A} = (A_1, A_2, A_3)$, respectively. The parameters $a_{Pi}$ and $a_{Ai}$, $\beta_{ij}$ and $\eta_{ij}$ are the "effective" Landau expansion coefficients; $g_{ijkl}$ is the polarization gradient tensor, $c_P$ and $c_A$ are the surface energy coefficients for $\vec{P}$ and $\vec{A}$, respectively; $\varepsilon_0$ is a universal dielectric constant, $\varepsilon_b$ is the relative background permittivity [33]. The polar order parameter $P_i$ is coupled with the electric field $E_i$. The magnitudes $P^2 = \sum_{i=1}^{3} P_i^2$, $A^2 = \sum_{i=1}^{3} A_i^2$ and $E^2 = \sum_{i=1}^{3} E_i^2$ are included in the $F_{el}$ and $F_S$.

Remarkably that the $x$-dependent "effective" Landau expansion coefficients [19, 32] are involved in Eq.(2a). The derivation of the expansion coefficients can be found in Refs.[19, 32] based on the experimental results of Park et. al. [5]. The effective coefficients describe the experimentally observed polar properties and phase diagrams of $Hf_xZr_{1-x}O_2$ thin films at room temperature. The coefficient extrapolation to the bulk material properties corresponds to the fact that bulk $HfO_2$ and $ZrO_2$ are high-k dielectrics without the ferroelectric soft mode and/or any other ferroelectric-like properties in a wide range of temperatures (below 1200 K) and pressures (below 12 GPa) [34].

The quasi-static electric field $E_i$ is related to the electric potential $\varphi$ as $E_i = -\frac{\partial \varphi}{\partial x_i}$. The potential $\varphi$ satisfies the Poisson equation inside the $Hf_xZr_{1-x}O_2$ nanoparticle and obeys the Debye-Huckell equation inside the ultra-thin screening shell with free charge carriers (mobile ions, electrons and/or vacancies) covering the nanoparticle (see **Fig. 1(a)**).

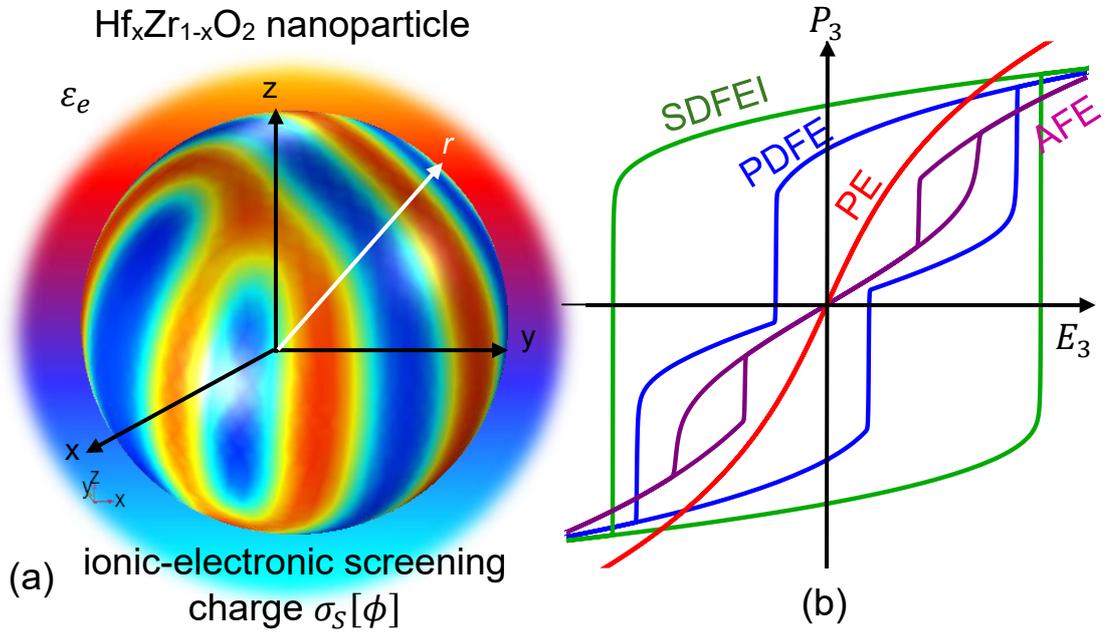

**FIGURE 1**. **(a)** The spherical $Hf_xZr_{1-x}O_2$ nanoparticle covered with the shell of ionic-electronic screening charges with the surface charge density $\sigma_S[\phi]$ adsorbed from the electrochemically active environment (or grain boundary) with free ionic-electronic charges. The relative dielectric constant of environment (or the effective dielectric permittivity of the inter-grain space) is $\varepsilon_e$. **(b)** Typical polarization response to external electric field in different phases and/or states of the $Hf_xZr_{1-x}O_2$ nanoparticles. The shown dependences $P_3(E_3)$ correspond to the paraelectric (PE) phase, antiferroelectric (AFE) state, poly-domain ferroelectric (PDFE) state and single-domain ferroelectric (SDFE) state.

Within the SH model the dependence of the surface charge density $\sigma_S[\phi]$ on electric potential excess $\delta\phi$ at the surface of the nanoparticle (or at the grain boundary) is controlled by the concentration of positive and negative surface charges $\theta_i[\phi]$ in a self-consistent manner. The charge density $\sigma_S[\phi]$ obeys the Langmuir adsorption isotherm [35, 36]:

$$\sigma_S[\phi] = \sum_i \frac{eZ_i\theta_i[\phi]}{N_i} \cong \sum_i eZ_i\, n_i \left(1 + \exp\left[\frac{\Delta G_i + eZ_i\delta\phi}{k_B T}\right]\right)^{-1}, \qquad (3)$$

where $e$ is the electron charge, $Z_i$ is the ionization number of the surface ions/vacancies, $1/N_i$ are the densities of positive and negative charge species in saturation ($i$=1,2), $n_i = 1/N_i$ is the surface charge density, $\Delta G_i$ are the formation energies of the surface charges (e.g., ions and/or vacancies) at normal conditions, $k_B$ is the Boltzmann constant and $T$ is the absolute temperature.

The linearization of Eq.(3) is possible for small built-in potentials, $\left|\frac{eZ_i\delta\phi}{k_B T}\right| < 1$. It leads to the approximate expression for the effective surface charge density $\sigma_S$, which is determined by the $\delta\phi$ and "effective" screening length $\lambda$ [36] as:

$$\sigma_S[\phi] \approx -\varepsilon_0 \frac{\delta\phi}{\lambda}, \qquad \frac{1}{\lambda} \approx \sum_i \frac{(eZ_i)^2 n_i}{4\varepsilon_0 k_B T \cosh^2\left(\frac{\Delta G_i}{2k_B T}\right)}. \qquad (4)$$

In the numerical calculations below we will consider the typical case [21, 22], when the surface charge formation energies are equal and relatively small, i.e., $\Delta G_1 = \Delta G_2 \cong 0.1$ eV, the ionization numbers are opposite, $Z_1 = -Z_2 = 1$, and vary the charge density $n_i$ from $10^{16}$m$^{-2}$ to $10^{20}$m$^{-2}$.

The minimization of the free energy $F$ allowing for the polarization relaxation leads to the coupled time-dependent Euler-Lagrange equations for the spatial-temporal evolution of the polar and antipolar order parameters, $P_i$ and $A_i$:

$$\Gamma_P \frac{\partial P_i}{\partial t} + a_{Pi}P_i + \beta_{ij}P_iP_j^2 + \eta_{ij}A_i^2 P_i - g_{ijkl}\frac{\partial P_k}{\partial x_j \partial x_l} = E_i, \qquad (4a)$$

$$\Gamma_A \frac{\partial A_i}{\partial t} + a_{Ai}A_i + \beta_{ij}A_iA_j^2 + \eta_{ij}P_i^2 A_i - g_{ijkl}\frac{\partial A_k}{\partial x_j \partial x_l} = 0. \qquad (4b)$$

Here $\Gamma_P$ and $\Gamma_A$ are Landau-Khalatnikov relaxation coefficients [37], and the summation over $i$ is absent. Typical boundary conditions to Eqs.(4) are of the third kind:

$$\left(c_P P_i - g_{ijkl}e_j \frac{\partial P_k}{\partial x_l}\right)\bigg|_S = 0, \quad \left(c_A A_i - g_{ijkl}e_j \frac{\partial A_k}{\partial x_l}\right)\bigg|_S = 0. \qquad (5)$$

Here $\vec{e}$ is the outer normal to the nanoparticle surface $S$. Sometimes it is convenient to introduce the extrapolation lengths [38], $\Lambda_P = \frac{g}{c_P}$ and $\Lambda_A = \frac{g}{c_A}$, where $g \equiv g_{44}$.

**B. Analytical description of the single-domain state of the Hf$_x$Zr$_{1-x}$O$_2$ nanoparticle**

The nanoscale Hf$_x$Zr$_{1-x}$O$_2$ can be a uniaxial ferroelectric (or antiferroelectric) depending on the Hf content $x$, size and screening conditions. Here we consider a single-domain Hf$_x$Zr$_{1-x}$O$_2$ nanoparticle, in which the direction of the spontaneous order parameters $\vec{P}_S$ and $\vec{A}_S$ are co-linear with the axis 3. For the geometry the electric field component $E_3$ is a superposition of external and depolarization fields, $E_3^e$ and $E_3^d$, respectively. For the spherical single-domain nanoparticle the analytical expressions for the electric field components have the form [39]:

$$E_3^d = -\frac{1}{\varepsilon_b + 2\varepsilon_e + (R/\lambda)} \frac{P_3}{\varepsilon_0}, \qquad E_3^e = \frac{3\varepsilon_e}{\varepsilon_b + 2\varepsilon_e + (R/\lambda)} E_3^0, \qquad (6)$$

where $R$ is the radius of the nanoparticle, $\lambda$ is the "effective" screening length corresponding to the ionic-electronic charge density; $\varepsilon_e$ is the relative effective dielectric constant of the environment (see **Fig. 1(b)**).

When $\lambda$ is very small (e.g., well below 0.1 nm), the ionic-electronic charge provides very good screening of the $Hf_xZr_{1-x}O_2$ spontaneous polarization and thus prevents the domain formation. For $\lambda \geq 0.1$ nm one should use the finite element modelling (FEM) [40, 41] or the phase-field approach [42] to account for the possible domain formation in the nanoparticle. The absence of the sixth order terms in the Landau energy (2a) and flexoelectric terms in the gradient energy (2b) allows the analytical description of the phase diagrams in the single-domain $Hf_xZr_{1-x}O_2$ nanoparticles. The ferroelectric domains morphology in the $Hf_xZr_{1-x}O_2$ nanoparticles will be calculated numerically by the FEM.

The depolarization field influence (given by Eqs.(6)), the spatial confinement effect (which can be responsible for the polar orthorhombic phase stability in nanoscale hafnia [1, 5, 9]) and polarization correlation effects lead to the following renormalization of the coefficient $a_P(x)$ and $a_A(x)$ in Eqs.(4):

$$\alpha_{PR}(x,R) = a_P(T,x)\left(\frac{R_P}{R}\right)^\gamma + \frac{\varepsilon_0^{-1}}{\varepsilon_b + 2\varepsilon_e + (R/\lambda)} + \frac{2g}{R\Lambda_P + R^2/4}, \qquad (7a)$$

$$\alpha_{AR}(x,R) = a_A(T,x)\left(\frac{R_A}{R}\right)^\gamma + \frac{2g}{R\Lambda_A + R^2/4}. \qquad (7b)$$

Hereinafter the effective sizes $R_P$ and $R_A$, and the positive factor $1 \leq \gamma \leq 2$ can be determined from the experiment considering the gradual disappearance of long-range polar and antipolar ordering in $Hf_xZr_{1-x}O_2$ films under their thickness increase, which was observed by e.g., Park et.al.[5] for the film thickness increase from 9 nm to 29 nm. The "net" coefficients $a_P(x)$ and $a_A(x)$ have been determined earlier [32, 20] using the experimental results of Park et. al. [5]. In accordance to the Gibbs model, the factors $\frac{R_P}{R}$ and $\frac{R_A}{R}$ may appear in Eqs.(7) from the ratio of the surface-induced orthorhombic phase Gibbs energy $G_O = 4\pi R^2 g_O$ to the bulk-induced monoclinic phase Gibbs energy $G_O = \frac{4\pi}{3} R^3 g_M$. The ratio is proportional to $\frac{1}{R}$ for a spherical particle. Using the surface bond contraction model, proposed by Huang et al [43], the factors $\left(\frac{R_P}{R}\right)^2$ and $\left(\frac{R_A}{R}\right)^2$ may appear in Eqs.(7). The term $\frac{\varepsilon_0^{-1}}{\varepsilon_b + 2\varepsilon_e + (R/\lambda)}$ in Eq.(7a) originates from the depolarization field, and the terms $\frac{2g}{R\Lambda_P+R^2/4}$ and $\frac{2g}{R\Lambda_A+R^2/4}$ in Eqs.(7a) and (7b) originate from the polarization correlation effects. The Landau expansion coefficients $a_P$ and $a_A$ are given in **Table A1** in Supplementary materials [44].

### C. Analytical description of the poly-domain state of the Hf$_x$Zr$_{1-x}$O$_2$ nanoparticle

Next, we consider the possibility of the domain structure appearance. Based on the results of Chenskii and Tarasenko for the "soft" domain structure onset [45] and using the direct variational method, the modified form of the LGDK-SH free energy density is:

$$f \approx \alpha_P \frac{P_3^2}{2} + \alpha_A \frac{A_3^2}{2} + \alpha_W \frac{W^2}{2} + b_P \frac{P_3^4}{4} + b_A \frac{A_3^4}{4} + b_W \frac{W^4}{4} + \eta_W \frac{P_3^2 W^2}{2} + \frac{\eta}{2}(P_3^2 + W^2)A_3^2 - E_3^e P_3. \quad (8)$$

Hereinafter $W$ is an additional order parameter, which is the spatial amplitude of the polarization periodic modulation, e.g., $P_3(x) \approx W sin(kx + \delta)$ (see **Table A2** and **Appendix B** in Ref.[44] for details). The renormalized expansion coefficients are:

$$\alpha_P = a_P \left(\frac{R_P}{R}\right)^\gamma + \frac{\varepsilon_0^{-1}}{\varepsilon_b + 2\varepsilon_e + (R/\lambda)} + \frac{2g}{R\Lambda_P+R^2/4}, \quad \alpha_A = a_A \left(\frac{R_A}{R}\right)^\gamma + \frac{2g}{R\Lambda_A+R^2/4}, \quad (9a)$$

$$\alpha_W \approx a_P \left(\frac{R_P}{R}\right)^\gamma + gk^2 + \frac{\varepsilon_0^{-1}}{(\varepsilon_b+2\varepsilon_e)[1+(\xi kR)^2]+(R/\lambda)}, \quad (9b)$$

$$b_P = b_A = \beta_{33}, \quad \eta_W = 3b_P, \quad b_W = \frac{9}{4}b_P, \quad \eta = \eta_{33}, \quad g = g_{44}. \quad (9c)$$

Here $\xi \approx 2/\pi$ is the geometrical factor and $k$ is the equilibrium value of the domain structure wave vector:

$$k \approx \frac{1}{\xi R}\sqrt{\frac{\xi R}{\sqrt{\varepsilon_0(\varepsilon_b+2\varepsilon_e)g}} - 1 - \frac{R}{(\varepsilon_b+2\varepsilon_e)\lambda}}. \quad (10)$$

The period of the domain structure is $\frac{2\pi}{k}$. The relative dielectric permittivity is given by expression:

$$\varepsilon = \varepsilon_b + \frac{1}{\varepsilon_0}\frac{\partial P_3}{\partial E_3^e} \xrightarrow[E_3^e \to 0]{} \begin{cases} \varepsilon_b + \frac{1}{\varepsilon_0 \alpha_P}, & \text{PE phase}, \\ \varepsilon_b + \frac{1/\varepsilon_0}{\alpha_P+3b_P P_3^2}, & \text{SDFE state}, \\ \varepsilon_b + \frac{1/\varepsilon_0}{\alpha_P+\eta_W W^2}, & \text{PDFE state}, \\ \varepsilon_b + \frac{1/\varepsilon_0}{\alpha_P+\eta A_3^2}, & \text{AFE phase}. \end{cases} \quad (11)$$

The phases and states abbreviations in Eq.(11) are described in **Fig. 1(b)**.

Expressions (9)-(11) have physical sense for $\frac{\xi(\varepsilon_b+2\varepsilon_e)}{\sqrt{\varepsilon_0(\varepsilon_b+2\varepsilon_e)g_{44}}} > \frac{\varepsilon_b+2\varepsilon_e}{R} + \frac{1}{\lambda}$, otherwise the polydomain state is unstable and the domain order parameter $W$ does not exist. The influence of finite size effect and ferro-ionic coupling is determined by the ratio $R/\lambda$, and the role of the environment is determined by the terms proportional to its dielectric constant $\varepsilon_e$.

The free energy (8) with the renormalized coefficients allows analytical description of the size effects, environment and ferro-ionic coupling influence on the phase diagrams and polar properties of the single- and poly-domain $Hf_xZr_{1-x}O_2$ nanoparticles. Corresponding analytical expressions for the energies of different phases, spontaneous polarization $P_S$, domain polarization amplitude $W_S$ and antipolar long-range order $A_S$ are listed in **Table A2** [44].

### III. RESULTS AND DISCUSSION
**A. The influence of size effects on the phase diagrams of $Hf_xZr_{1-x}O_2$ nanoparticles**

Phase diagram of the spherical $Hf_xZr_{1-x}O_2$ nanoparticle, calculated in dependence of the Hf content $x$ and the particle radius $R$, is shown in **Fig. 2(a)**. The diagram is calculated for typical value of the surface charge density $n_i = 10^{17} m^{-2}$ and high-k environment with $\varepsilon_e = 10$. The color scale in the figure shows the magnitudes of the domain structure polarization $W_S$ and antipolar order $A_S$ in $\mu C/cm^2$. Almost straight horizontal lines are the boundaries between the paraelectric (PE) phase, where $P_S = A_S = W_S = 0$ (the yellow region located at $0.71 < x < 1$); the poly-domain ferroelectric (PDFE) state, where $W_S > 0$ and $P_S = A_S = 0$ (the reddish region located at $0.49 < x < 0.7$); the antiferroelectric (AFE) phase, where $A_S < 0$ and $P_S = W_S = 0$ (the bluish region located at $0.1 < x < 0.49$); and the dielectric (DE) state without any long-range order (the yellow region located at $0 \leq x < 0.1$). Similarly to the case of $Hf_xZr_{1-x}O_2$ thin films, the increase in $x$ from 0 to 1 induces the series of phase transitions DE→AFE→PDFE→PE. Also, we observed the gradual and monotonic decrease of $W_S$ and $A_S$ in $Hf_xZr_{1-x}O_2$ nanoparticles under their size increase (see the color contrast decrease with increase in $R$). The apparent disappearance of the polar and antipolar long-range ordering corresponds to $R \gg R_{P,A}$ in accordance with Eqs.(9).

The dependences of the static dielectric permittivity $\varepsilon$ on the Hf content $x$ calculated in zero electric field ($E_3^e \to 0$) for several values of $R$ from 2 nm to 10 nm are shown in **Fig. 2(b)**. The fractures of the permittivity curves (marked by the arrows) are associated with the first order phase transitions from the DE state to the AFE phase (at $x \approx 0.1$), from the AFE phase to the PDFE state (at $x \approx 0.5$), and from the PDFE state to the PE phase (at $x \approx 0.7$). Since the strength $\eta$ of the biquadratic coupling term in Eq.(8) is relatively high (see **Table A1** in Ref.[44]), the appearance and changes of the "hidden" antipolar order significantly change the "observable" dielectric permittivity $\varepsilon$. The dielectric permittivity monotonically increases with increase in $R$ due to the gradual decrease and eventual disappearance of the polar and/or antipolar ordering.

To summarize, the influence of size effect on the phase diagrams, polar and antipolar long-range orders, and dielectric permittivity is very strong for the small radii (1 nm $\leq R \leq$ 10 nm) of $Hf_xZr_{1-x}O_2$ nanoparticles.

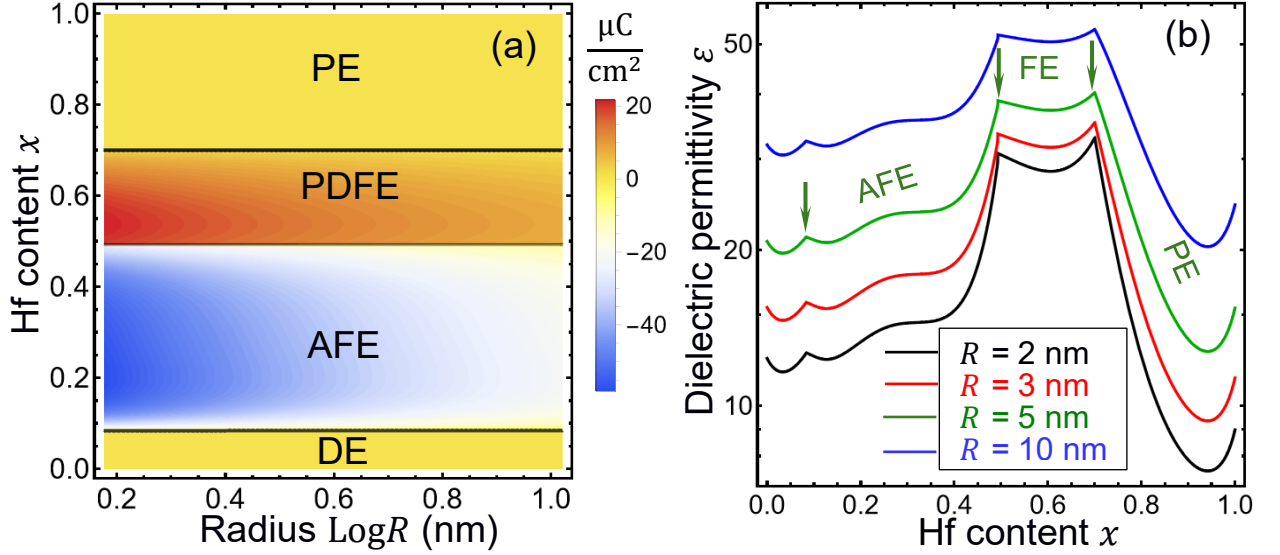

**FIGURE 2. (a)** The phase diagram of the $Hf_xZr_{1-x}O_2$ nanoparticle in dependence on the Hf content $x$ and particle radius $R$. Color scale shows the values of the order parameters $P_S$, $W_S$ and $A_S$ in µC/cm². **(b)** The dependence of the nanoparticle dielectric permittivity $\varepsilon$ on the Hf content $x$ calculated for $R$ =2 nm (the black curve), 3 nm (the red curve), 5 nm (the green curve) and 10 nm (the blue curve). The diagrams are calculated for the following parameters: $\varepsilon_b$ =3, $\Delta G_i$ =0.1 eV, $Z_1 = -Z_2 = 1$, $n_i = 10^{17} m^{-2}$ and $\varepsilon_e$ =10.

### B. The influence of the environment on the phase state of $Hf_xZr_{1-x}O_2$ nanoparticles

Phase diagram of the $Hf_xZr_{1-x}O_2$ nanoparticle, calculated in dependence of the Hf content $x$ and the environment relative dielectric constant $\varepsilon_e$, is shown in **Fig. 3(a)**. The diagram is calculated for the surface charge density $n_i = 10^{17} m^{-2}$ and the particle radius $R = 5$ nm. The color scale in the figure shows the magnitudes of the spontaneous polarization $P_S$, domain structure polarization amplitude $W_S$ and antipolar order $A_S$ in $\frac{\mu C}{cm^2}$. Almost straight horizontal lines separate the PE phase (the yellow region located at $0.71 < x < 1$); the PDFE state (the reddish region located at $0.49 < x < 0.71$); the AFE phase (the bluish region located at $0.1 < x < 0.49$); and the DE state (the yellow region located at $0 \leq x < 0.1$). The polarization amplitude $W_S$ relatively weakly depends on $\varepsilon_e$ in the PDFE state. The reentrant single-domain ferroelectric (SDFE) state, where $P_S > 0$ and $W_S = A_S = 0$, occupies the red-colored U-shaped region located inside the PDFE state, where the environment dielectric constant is high $\varepsilon_e > \varepsilon_{cr}$ ($\varepsilon_{cr} \approx 150$ for chosen material parameters) and $0.52 < x < 0.67$. The high value of $\varepsilon_e$ is required to reduce the depolarization field factor in Eqs.(7) and (9), and hence to suppress the domain formation.

The dependences of the static dielectric permittivity $\varepsilon$ on the Hf content $x$ calculated at $E_3^e \to 0$ are shown in **Fig. 3(b)** for several values of the nanoparticle radius from 2 nm to 10 nm. Three fractures at the permittivity curves (marked by green arrows) correspond to the first order phase transitions from the DE state to the AFE phase (at $x \approx 0.1$), from the AFE phase to the PDFE state (at $x \approx 0.5$), and from the PDFE state to the PE phase (at $x \approx 0.7$). These three successive transitions are observed for $\varepsilon_e < \varepsilon_{cr}$ (see the black, red and green permittivity curves calculated at $\varepsilon_e = 1$, 10 and 81). For $\varepsilon_e > \varepsilon_{cr}$ the five first order phase transitions correspond to the fractures on the permittivity curves (see the blue arrows point near the permittivity curve calculated at $\varepsilon_e = 300$). Two additional fractures correspond to the PDFE→SDFE and SDFE→PDFE transitions. The dielectric permittivity monotonically increases with increase in $\varepsilon_e$ until the SDFE state appears for high $\varepsilon_e$. The increase is related with the gradual decrease of the depolarization energy caused by the increase in $\varepsilon_e$ that eventually facilitates the appearance of the polar long-range order. Due to the spontaneous polarization appearance at $\varepsilon_e > \varepsilon_{cr}$, the permittivity in the SDFE state is smaller than the permittivity in the PDFE state.

Thus, the influence of the environment on the polar long-range order becomes significant for $\varepsilon_e > \varepsilon_{cr}$, where the critical value $\varepsilon_{cr}$ is defined by material properties, screening conditions and size effects. Meanwhile the influence of $\varepsilon_e$ on the antipolar order is absent due to the absence of the depolarization field analog and polar-type domain walls in the uniaxial antiferroelectrics.

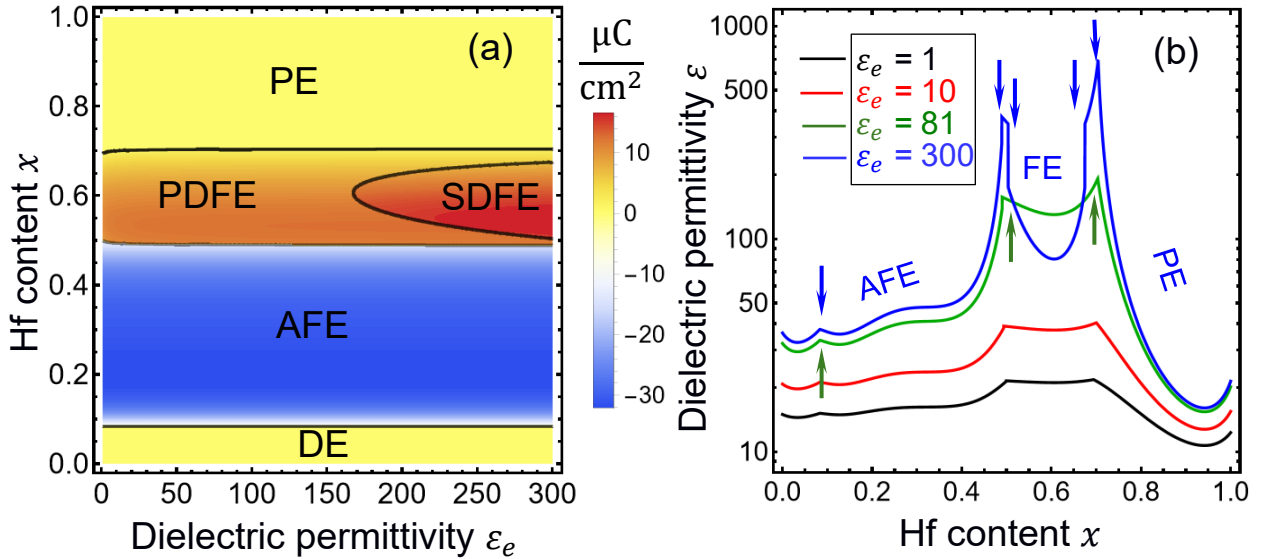

**FIGURE 3.** The phase diagram of the $Hf_xZr_{1-x}O_2$ nanoparticle in dependence on the Hf content $x$ and environment dielectric constant $\varepsilon_e$. Color scale shows the values of the order parameters $P_S$, $W_S$ and $A_S$ in µC/cm². **(b)** The dependence of the nanoparticle dielectric permittivity $\varepsilon$ on the Hf content $x$ calculated for $\varepsilon_e = 1$ (the black curve), 10 (the red curve), 81 (the green curve) and 300 (the blue curve). The particle radius is $R = 5$ nm; $n_i = 10^{17}$m$^{-2}$; other parameters are the same as in **Fig. 2**.

### C. The ferro-ionic states in the Hf$_x$Zr$_{1-x}$O$_2$ nanoparticles

Phase diagram of the Hf$_x$Zr$_{1-x}$O$_2$ nanoparticle, calculated in dependence of the Hf content $x$ and the ionic-electronic charge density $n_i$, is shown in **Fig. 4(a)**. The diagram is calculated for $R = 5$ nm and $\varepsilon_e = 10$. The structure of the diagram and the distribution of colors are similar to **Fig. 3(a)**, because both high $\varepsilon_e$ and high $n_i$ decrease strongly the depolarization field contribution (see e.g., the denominators in Eqs.(9) proportional to $2\varepsilon_e + (R/\lambda)$). However, here the single-domain ferro-ionic (SDFEI) state appears at high $n_i$. Black horizontal lines separate the PE phase (the yellow region located at $0.71 < x < 1$); the PDFE state (the reddish region located at $0.49 < x < 0.71$); the AFE phase (the bluish region located at $0.1 < x < 0.49$); and the DE state (the yellow region located at $0 \leq x < 0.1$) in the diagram. The domain polarization amplitude $W_S$ relatively weakly depends on $n_i$ inside the region of the PDFE state. The partially reentrant SDFEI state replaces the PDFE state at $n_i > n_{cr}$, where $n_{cr} \approx 2 \cdot 10^{18}$ m$^{-2}$ for chosen material parameters (see the red region at $0.5 < x < 0.7$). The high values of $n_i$ reduce the depolarization field factor in Eqs.(7) and (9), and simultaneously induce the ferro-ionic state via the overpotential mechanism.

The dependences of the static dielectric permittivity $\varepsilon$ on the Hf content $x$ calculated at $E_3^e \to 0$ are shown in **Fig. 4(b)** for several values of the $n_i$. The fractures at the permittivity curves (marked by the black arrows) originate due to the first order phase transitions from the DE state to the AFE phase (at $x \approx 0.1$), from the AFE phase to the PDFE state (at $x \approx 0.5$), and from the PDFE state to the PE phase (at $x \approx 0.7$). These three successive transitions are observed for $n_i < n_{cr}$ (see the black, red and green permittivity curves calculated at $n_i = 10^{17}$ m$^{-2}$, $3 \cdot 10^{17}$ m$^{-2}$ and $10^{18}$ m$^{-2}$). For $n_i > n_{cr}$ the five first order phase transitions correspond to the fractures at the permittivity curves (see the blue arrows near the permittivity curve calculated at $n_i = 3 \cdot 10^{18}$ m$^{-2}$). Two additional fractures correspond to the PDFE→SDFEI and SDFEI→PDFE transitions. The dielectric permittivity monotonically increases with increase in $n_i$ until the SDFEI state appears for $n_i$ higher than $n_{cr}$. The increase is related with the gradual appearance of the ferro-ionic long-range order. Due to the spontaneous polarization appearance at $n_i > n_{cr}$, the permittivity in the SDFEI state is smaller than the permittivity in the PDFE state.

Let us underline that the influence of the ionic-electronic charge density $n_i$ on the phase diagram and spontaneous polarization of Hf$_x$Zr$_{1-x}$O$_2$ nanoparticle is significant in the $x$-range $0.5 < x < 0.7$ corresponding to the polar state appearance, being the origin of the SDFEI state for $n_i > n_{cr}$. The influence of $n_i$ on the antipolar long-range order is absent, because the depolarization field is absent in the AFE and DE phases.

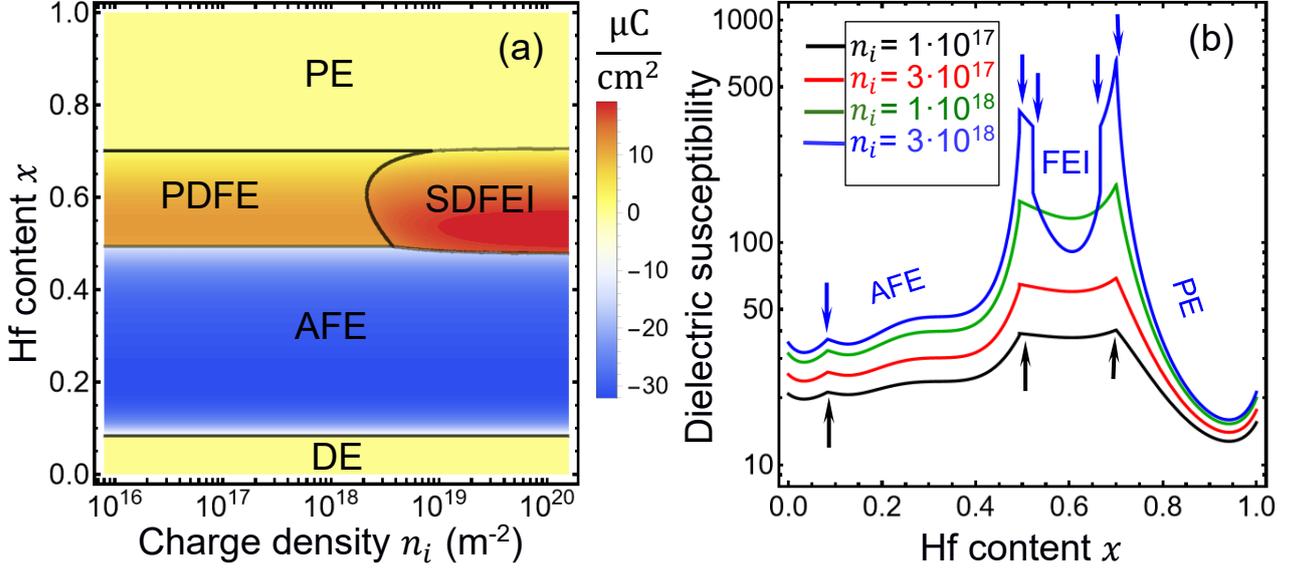

**FIGURE 4. (a)** The phase diagram of $Hf_xZr_{1-x}O_2$ nanoparticle in dependence on the Hf content $x$ and ionic-electronic charge density $n_i$. Color scale shows the values of the order parameters $P_S$, $W_S$ and $A_S$ in $\mu C/cm^2$. **(b)** The dependence of the nanoparticle dielectric susceptibility on the Hf content $x$ calculated for $n_i = 10^{17} m^{-2}$ (the black curve), $3 \cdot 10^{17} m^{-2}$ (the red curve), $10^{18} m^{-2}$ (the green curve) and $3 \cdot 10^{18} m^{-2}$ (the blue curve). The radius is $R = 5$ nm and $\varepsilon_e = 10$, other parameters are the same as in **Fig. 2.**

### D. The influence of the ferro-ionic coupling on the domain morphology in $Hf_xZr_{1-x}O_2$ nanoparticles

As it follows from the diagrams in **Figs. 2**, **3** and **4**, the influence of the surface ionic-electronic charge on the polar phases of $Hf_xZr_{1-x}O_2$ nanoparticles is more important in comparison with the size and/or environment changes. However, the diagrams do not give us any quantitative information about the influence of these parameters on the domain morphology in the nanoparticles. Indeed, the relatively small screening charge density $n_i \leq n_{cr}$ and/or environment dielectric constant $\varepsilon_e \leq \varepsilon_{cr}$ cannot support the homogeneous spontaneous polarization inside the nanoparticle. In result the single-domain polar state has higher free energy in comparison with the poly-domain state, at that the domain morphology can be influenced strongly by the environment and ionic-electronic screening. Below we discuss typical FEM results, which illustrate the influence of size effect, environment dielectric properties and ionic-electronic screening on the domain morphology in $Hf_xZr_{1-x}O_2$ nanoparticles.

Typical changes of the domain structure morphology, which appear in $Hf_xZr_{1-x}O_2$ nanoparticles with increase in $\varepsilon_e$, are shown in **Figs. 5(a)-(d)** calculated for 10-nm $Hf_{0.6}Zr_{0.4}O_2$ nanoparticles and $n_i = 10^{17} m^{-2}$. For $\varepsilon_e < 10$ the fine and faint labyrinthine domains appear in the central part of the nanoparticle, while its subsurface layer is paraelectric (see **Fig. 5(a)**). The labyrinthine average period and contrast (i.e., the polarization amplitude) increase significantly with increase in $\varepsilon_e$ from 10 to 81 (see **Fig. 5(b)** and **5(c)**). The labyrinthine domains are multiple-degenerated metastable energy states, and the concrete maze depends on the initial polarization distribution in the random seeding. The main physical origin of the mazes' stability (in comparison with regular domain stripes) is the surface curvature effect that is significant in the small nanoparticles. For $\varepsilon_e > \varepsilon_{cr}$ the labyrinthine domains gradually vanish, being replaced by the regular structures, then the regular structures are replaced by the regions of the single-domain state at $\varepsilon_e \gg \varepsilon_{cr}$ (see **Fig. 5(d)**).

The effect of the "geometric catastrophe", which happens in ultra-small nanoparticles with increase in $\varepsilon_e$, are shown in **Figs. 5(e)-(h)** for 6-nm $Hf_{0.6}Zr_{0.4}O_2$ nanoparticles and $n_i = 10^{17} m^{-2}$. Here the labyrinthine domains are almost absent in the studied range of $\varepsilon_e$ and the paraelectric state of the nanoparticle is stable for $\varepsilon_e < 9$. For $\varepsilon_e \geq 10$ the faint domain stripes appear in the central part of the nanoparticle (see **Fig. 5(e)** and **5(f)**). The stripes period and contrast increase strongly with increase in $\varepsilon_e$ above 50 (see **Fig. 5(g)**) and the gradual transition to the bi-domain and/or single-domain states appear for $\varepsilon_e \gg \varepsilon_{cr}$ (see **Fig. 5(h)**).

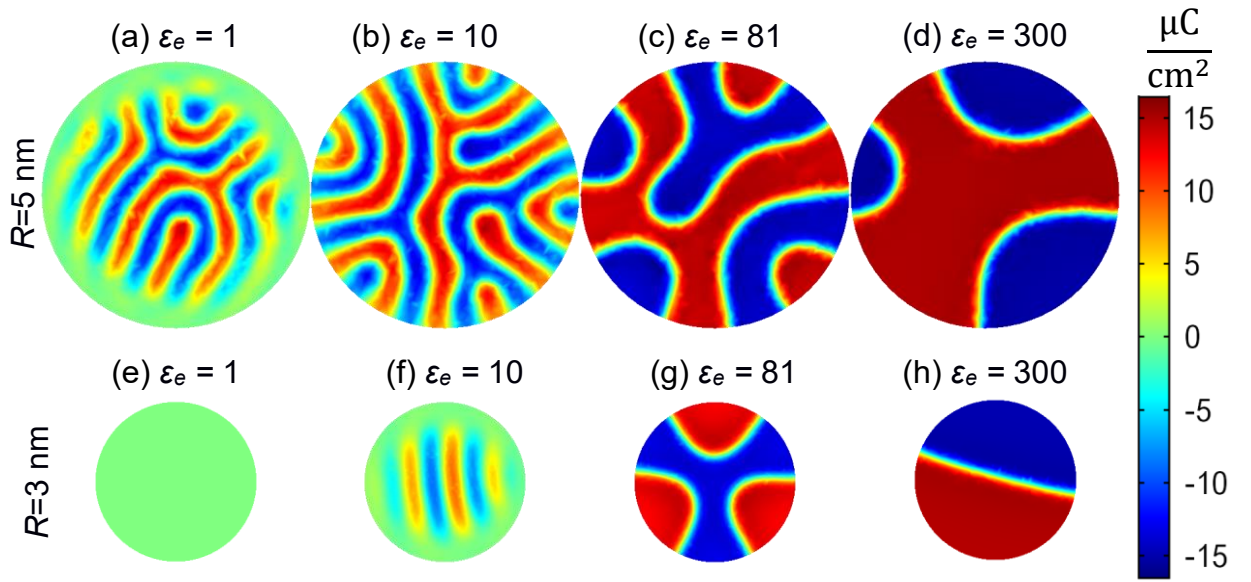

**FIGURE 5.** Typical view of the equilibrium ferroelectric domain structures calculated by the FEM in the $Hf_{0.6}Zr_{0.4}O_2$ nanoparticle covered by the ionic-electronic screening charges with the surface density $n_i = 10^{17} m^{-2}$. Particles radius $R = 5$ nm (**a - d**) or 3 nm (**e - h**). The relative dielectric constant of environment

is $\varepsilon_e = 10$ **(a, e)**, 10 **(b, f)**, 81 **(c, g)** and 300 **(d, h)**. other parameters are the same as in **Fig. 2**. Color scale shows the polarization value in µC/cm².

Typical changes of the domain structure morphology, which appear in $Hf_xZr_{1-x}O_2$ nanoparticles with increase in $n_i$, are shown in **Figs. 6(a)-(d)** calculated for 10-nm $Hf_{0.6}Zr_{0.4}O_2$ nanoparticles and $\varepsilon_e = 10$. For $3 \cdot 10^{16} m^{-2} \leq n_i \leq 3 \cdot 10^{17} m^{-2}$ the fine labyrinthine domains appear inside the nanoparticle (see **Fig. 6(a)** and **6(b)**). The average period and contrast of the mazes increase with increase in $n_i$ (see **Fig. 6(c)**). For $n_i \geq n_{cr}$ the gradual transition to the single-domain ferro-ionic state occurs (see **Fig. 6(d)**).

The geometric catastrophe, leading to the maze disappearance or their strong suppression, is evident from **Figs. 6(e)-(h)** calculated for 6-nm $Hf_{0.6}Zr_{0.4}O_2$ nanoparticles and $\varepsilon_e = 10$. For relatively small $n_i$ (namely $10^{17} m^{-2} \leq n_i \leq 3 \cdot 10^{17} m^{-2}$) the labyrinthine domains are replaced by faint stripe domains, which exist in the central part of the nanoparticle. For $n_i \geq 10^{18} m^{-2}$ the faint labyrinthine domains appear and fill the whole nanoparticle (see **Fig. 6(g)**). The gradual transition to the bi-domain and/or single-domain ferro-ionic states appear for $n_i \geq 3 \cdot 10^{18} m^{-2}$ (see **Fig. 6(h)**). From comparison of **Fig. 5** and **6**, as well as from the analysis of the FEM results performed for wider range of sizes, environment dielectric constant and ionic-electronic charge density, we concluded that the ferro-ionic coupling, which is significant for relatively high $n_i$, not only support the polar long-range order (due to the increase of screening degree) and induce the single-domain ferro-ionic state, but also can induce and/or enlarge the region of the labyrinthine domains stability towards smaller radii $R$ and smaller dielectric constants $\varepsilon_e$.

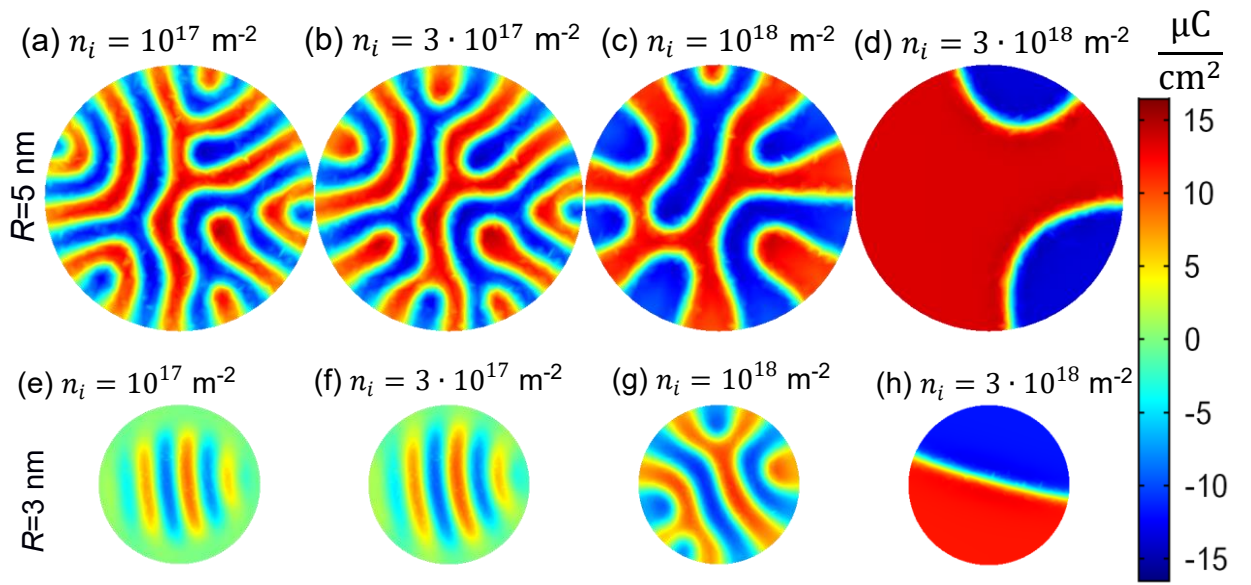

**FIGURE 6.** Typical view of the equilibrium ferroelectric domain structures calculated by the FEM in the Hf$_{0.6}$Zr$_{0.4}$O$_2$ nanoparticle covered by the ionic-electronic screening charges with the surface density $n_i = 10^{17}$ m$^{-2}$ **(a, e)**, $3 \cdot 10^{17}$ m$^{-2}$ **(b, f)**, $10^{18}$ m$^{-2}$ **(c, g)** and $3 \cdot 10^{18}$ m$^{-2}$ **(d, h)**. Particles radius $R = 5$ nm **(a - d)** or 3 nm **(e - h)**. The relative dielectric constant of environment is $\varepsilon_e = 10$; other parameters are the same as in **Fig. 2**. Color scale shows the polarization value in µC/cm$^2$.

In the next subsection we show that the appearance multiple-degenerated domain states (such as mazes) can significantly influence the negative differential capacitance state in the nanograined Hf$_{0.6}$Zr$_{0.4}$O$_2$ films, which grains are covered by the screening charge shells.

### E. The negative capacitance state in the nanograined Hf$_x$Zr$_{1-x}$O$_2$ films

The differential negative capacitance (NC) state is successfully used in advanced nanoelectronics [46]. Replacement of the positive capacitance insulator film in the gate stack of a field-effect transistor (FET) with a thin Si-compatible ferroelectric film in the NC state has several evident advantages, such as the significant decrease of the stack heating and significant reduction of the subthreshold swing of the FETs below the thermodynamic Boltzmann limit [47, 48, 49]. The stabilization of the NC state was revealed experimentally a while ago in the double-layer capacitor made of the paraelectric and ferroelectric layers, which total capacitance appeared greater than it would be for the dielectric layer of the same thickness as used in the double-layer capacitor [50, 51].

Today many experimental demonstrations of the stable NC state in ferroelectric double-layer capacitors exist [52, 53, 54, 55]. The general-accepted point of view is that the static NC effect can be pronounced in the vicinity of the boundary between the paraelectric and ferroelectric phases [52]. At that the influence of the domain structure emergence in the ferroelectric layer on the NC state can be very significant (see e.g., theoretical papers [56, 57, 58]). However, to the best of our knowledge, is it still difficult to find the conditions of the NC state emergence and stability considering the inevitable domain formation in the nano-grained ferroelectric films. As we have shown above, the domain morphology inside the separated grains can be very complex, e.g., the labyrinthine domains can appear due to the surface curvature effect.

Our analytical calculations and FEM show that the nanograined Hf$_x$Zr$_{1-x}$O$_2$ film, which surface is separated from the electrode by a thin paraelectric layer, can be in the NC state emerging from the Hf$_x$Zr$_{1-x}$O$_2$ nanograins covered by the screening charge shells (see **Fig. 7(a)**). The expression for the relative differential capacitance of the double-layer capacitor with the ionic-electronic surface charge is [31]:

$$\frac{\Delta C}{C_r} = \frac{C_r - C_{eff}}{C_r} = \frac{h_f \varepsilon_e - d\varepsilon_b}{h_e \varepsilon_b + h_f \varepsilon_e} \frac{d\sigma}{dU} + \frac{2h_f \varepsilon_e}{h_e \varepsilon_b + h_f \varepsilon_e} \frac{d\bar{P}}{dU} - 2\frac{\varepsilon_0 \varepsilon_b \varepsilon_e}{h_e \varepsilon_b + h_f \varepsilon_e} - \frac{\varepsilon_0 \varepsilon_e}{h_e}, \quad (12)$$

where the "reference" capacitance of the single-layer dielectric capacitor is $C_r = \frac{\varepsilon_0 \varepsilon_e}{h_e}$, $C_{eff}$ is the effective capacitance of the double-layer capacitor, $U$ is the voltage applied to the double-layer capacitor, $h_f$ is the thickness of the Hf$_x$Zr$_{1-x}$O$_2$ film, $h_e$ is the thickness of the paraelectric layer with the relative dielectric permittivity $\varepsilon_e$. $\bar{P}$ is the average polarization of the film and $\sigma$ is the average density of the ionic-electronic surface charge. Using the approach proposed in Ref.[59], we derived that the NC effect can be reached under the conditions:

$$-\frac{2h_e}{\varepsilon_e h_f + 2\varepsilon_b h_e} < \frac{\alpha_A + 3b_A A_3^2 + \eta(P_3^2 + W^2)}{\left(\alpha_f + 3b_P P_3^2 + \eta A_3^2 + \eta_W W^2\right)\left(\alpha_A + 3b_A A_3^2 + \eta(P_3^2 + W^2)\right) - 4\eta^2 P_3^2 A_3^2} < 0, \quad (13)$$

where $\alpha_f = a_P \left(\frac{R_P}{R}\right)^\gamma + \frac{2g}{R\Lambda_P + R^2/4}$, and parameters $\alpha_A$, $b_A$, $b_P$, $\eta$ and $\eta_W$ are given by Eqs.(9).

The NC state in dependence on the Hf content $x$ and ionic-electronic charge density $n_i$ calculated considering the domain appearance is located inside the small region between solid curves in **Fig. 7(b)**. the NC state calculated without domains is located inside the much larger region between dashed curves in **Fig. 7(b)**. The strong shrinking of the NC region area is explained by the circumstance that the PDFE state replaces the PE phase, where the NC state may appear in the single-domain approximation. These results confirm that the total effect of a static domain structure is to reduce the stability range of the static NC [52, 60].

However, we obtained that the multiple-degenerated domain states (such as mazes) significantly facilitate the low-frequency dynamic NC effect in the spherical Hf$_x$Zr$_{1-x}$O$_2$ nanoparticles, as it was shown earlier for spherical Sn$_2$P$_2$S$_6$ nanoparticles [61]. The region of the dynamic NC effect fills the most part of the area between dashed curves in **Fig. 7(b)**.

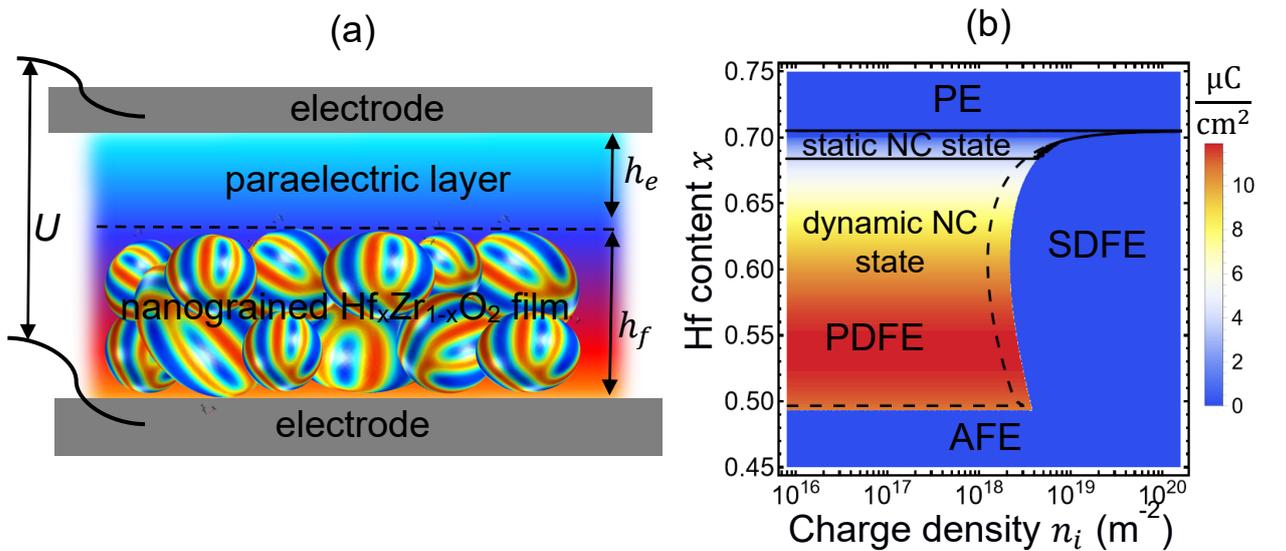

**FIGURE 7.** (a) The double-layer capacitor consisting of the nanograined Hf$_x$Zr$_{1-x}$O$_2$ film and the paraelectric layer. **(b)** The NC state in dependence on the Hf content $x$ and ionic-electronic charge density $n_i$ calculated considering the domain appearance is located inside the small region between the solid curves; the NC state calculate without domains is located inside the much larger region between the dashed curves. Color scale shows the values of the order parameter $W$ in µC/cm². The radius is $R = 5$ nm and $\varepsilon_e = 10$, other parameters are the same as in **Fig. 2**.

## IV. CONCLUSIONS

We studied the influence of content $x$, size effects, environment dielectric constant and concentration of surface ions on the phase diagrams, dielectric permittivity, spontaneous polar and antipolar ordering, and domain structure morphology in spherical $Hf_xZr_{1-x}O_2$ nanoparticles, which surface interacts with electrochemically active environment.

Appeared that the influence of size effect on the phase diagrams, polar and dielectric properties, and domain structure morphology is very strong for the small radii (1 nm≤ $R$ ≤10 nm) of $Hf_xZr_{1-x}O_2$ nanoparticles. The influence of the environment dielectric constant $\varepsilon_e$ on the polar long-range order becomes significant for $\varepsilon_e$ above the critical value $\varepsilon_{cr}$ defined by material properties, screening conditions and size effects. The influence of $\varepsilon_e$ on the antipolar order is absent due to the absence of the depolarization field analog and polar-type domain walls in the uniaxial antiferroelectrics.

The influence of the ionic-electronic charge density $n_i$ on the phase diagram and spontaneous polarization of $Hf_xZr_{1-x}O_2$ nanoparticle is significant in the x-range 0.5≤x≤0.7 corresponding to the polar state appearance, being the origin of the SDFEI state for $n_i > n_{cr}$, where the critical value $n_{cr}$ is defined by material properties, environment dielectric properties and size effects. The influence of $n_i$ on the antipolar long-range order is absent, because the depolarization field is absent in the antiferroelectric and dielectric phases.

Not less important is that the ferro-ionic coupling supports the polar long-range order for 0.49≤x≤0.71, induces and/or enlarges the stability region of the labyrinthine domains towards smaller sizes and smaller environmental dielectric permittivity at low concentrations of the surface ions $n_i \ll n_{cr}$, as well as causes the transition to the single-domain ferro-ionic state at high concentrations of the surface ions $n_i \geq n_{cr}$. We also predict that the labyrinthine domain states, being multiple-degenerated, significantly influence the appearance of the differential NC state in the nanograined/nanocrystalline $Hf_xZr_{1-x}O_2$ films.

Obtained results are applicable for the polycrystalline $Hf_xZr_{1-x}O_2$ films, which can be imagined as an ensemble of single-crystalline nanoparticles separated by grain boundaries capturing screening charges (such as charged vacancies and/or ions) and having finite densities of electronic and ionic states.

**Acknowledgements.** The work of A.N.M. and E.A.E. are funded by the National Research Foundation of Ukraine (projects "Manyfold-degenerated metastable states of spontaneous polarization in nanoferroics: theory, experiment and perspectives for digital nanoelectronics", grant N 2023.03/0132 and "Silicon-compatible ferroelectric nanocomposites for electronics and sensors", grant N 2023.03/0127). This effort (problem statement and general analysis, S.V.K.) was supported as part of the center for 3D Ferroelectric Microelectronics (3DFeM), an Energy Frontier Research Center funded by the U.S. Department of Energy (DOE), Office of Science, Basic Energy Sciences under Award Number DE-SC0021118. A.N.M. also acknowledges the Horizon Europe Framework Programme (HORIZON-TMA-MSCA-SE), project № 101131229, Piezoelectricity in 2D-materials: materials, modeling, and applications (PIEZO 2D). Numerical results presented in the work are obtained and visualized using a specialized software, Mathematica 14.0 (https://www.wolfram.com/mathematica).

**Authors' contribution.** A.N.M. and S.V.K. generated the research idea, formulated the problem, and wrote the manuscript draft. E.A.E. performed analytical calculations, wrote the codes and prepared corresponding figures. All co-authors discussed the results and worked on the manuscript improvement.

# SUPPLEMENTARY MATERIALS
## to
### "Ferro-ionic States and Domains Morphology in Hf$_x$Zr$_{1-x}$O$_2$ Nanoparticles"

by

Eugene A. Eliseev[1], Sergei V. Kalinin[2*], and Anna N. Morozovska[3†]

[1] Frantsevich Institute for Problems in Materials Science, National Academy of Sciences of Ukraine

Omeliana Pritsaka str., 3, Kyiv, 03142, Ukraine

[2] Institute of Physics, National Academy of Sciences of Ukraine,

pr. Nauky 46, 03028, Kyiv, Ukraine

[3] Department of Materials Science and Engineering, University of Tennessee,

Knoxville, TN, 37996, USA

**Appendix A. The parameters of the "effective" LGDK model for Hf$_x$Zr$_{1-x}$O$_2$ nanoparticles**

The x-dependences of the coefficients $a_P(x)$, $b_P(x)$, $b_A(x)$, $a_A(x)$, and $\eta(x)$ were determined from the fitting of experimental results of Park et al. [1] and further interpolated in the x-range ($0 \leq x \leq 1$) in Ref.[2, 3]. The polynomial x-functions, used for the dependence of the effective LGD expansion coefficients are listed in **Table AI**. The interpolations are valid in the narrow range of the nanoparticle sizes, 2 nm $\leq R \leq$ 20 nm, because smaller nanoparticles lose their ferroelectric properties due to the correlation effect and bigger nanoparticles should become dielectric due to the structural transition into the monoclinic phase (similarly to the case of thin films).

**Table AI.** The interpolated effective LGDK parameters of Hf$_x$Zr$_{1-x}$O$_2$ as the function of $x$

| coeff. | units | |
|---|---|---|
| $a_P(x)$ | $10^{10} \frac{m}{F}$ | $0.32266 + 2.6610\,x - 50.692\,x^2 + 239.60\,x^3 - 505.58\,x^4 + 493.35\,x^5 - 179.07\,x^6$ |
| $a_A(x)$ | $10^{10} \frac{m}{F}$ | $0.10924 - 1.5215\,x + 2.5362\,x^2$ |
| $b_P(x)$ | $10^{10} \frac{Vm^5}{C^3}$ | $1.4535 - 22.3815\,x + 174.68\,x^2 - 462.96\,x^3 + 405.86\,x^4$ |
| $b_A(x)$ | $10^{10} \frac{Vm^5}{C^3}$ | $b_A(x) \approx b_P(x)$ |

---

[*] Corresponding author: sergei2@utk.edu

[†] Corresponding author: anna.n.morozovska@gmail.com

| $\eta(x)$ | $10^{10}\frac{Vm^5}{C^3}$ | $0.60712 + 7.5685\,x - 4.6661\,x^2$ |

The effective free energy with renormalized coefficients is:

$$f[P,A,W] \approx \frac{a_R}{2}P^2 + \frac{a_A}{2}A^2 + \frac{a_W}{2}W^2 + \frac{b_P}{4}P^4 + \frac{b_A}{4}A^4 + \frac{b_W}{4}W^4 + \frac{\eta_W}{2}P^2W^2 + \frac{\eta}{2}(P^2 + W^2)A^2 \quad (A.1)$$

The magnitude of the spontaneous polarization $P_s$, antipolar order $A_s$ and domain amplitude $W_s$, corresponding free energy densities $f_D$, stability conditions of the thermodynamically stable phases of the free energy (A.1) and critical field(s) are listed in **Table A2**. The columns for the DE, PE, SDFE and AFE phase are the same as in Ref.[2] (as it should be).

**Table A2.** Thermodynamically stable phases of the free energy (A.1).

| Phase and type of hysteresis loops | Spontaneous values of the order parameters | Free energy density and stability conditions | Coercive and/or critical field(s) $E_c$, and/or $P(E_c)$ * |
|---|---|---|---|
| Dielectric (DE) and paraelectric (PE) phases. | $P_s = A_s = W_s = 0$ | $f_D = 0$<br>$a_R > 0, a_A > 0, a_W > 0$ | absent |
| Ferroelectric single-domain phase (SDFE) | $A_s = 0$,<br>$P_s = \pm\sqrt{-\frac{a_R}{b_P}}$,<br>$W_s = 0$ | $f_P = -\frac{a_R^2}{4b_P}$<br>$f_P = min,\ a_R < 0$,<br>$a_A b_P - \eta a_R > 0$<br>$a_W b_P - \eta_W a_R > 0$ | $E_c = \pm\frac{2}{3\sqrt{3}}\frac{(-a_R)^{3/2}}{b_P}$<br>$P_3(E_c) = 0$ |
| Mixed phase (FI) (which appears unstable for material parameters from **Table A1**) | $A_s = \pm\sqrt{-\frac{a_A b_P - \eta a_R}{b_A b_P - \eta^2}}$,<br>$P_s = \pm\sqrt{-\frac{a_R b_A - \eta a_A}{b_A b_P - \eta^2}}$,<br>$W_s = 0$ | $f_{PA} = \frac{-b_P a_A^2 - b_A a_R^2 + \eta a_P a_A}{4(b_A b_P - \eta^2)}$<br>$f_{PA} = min$,<br>$a_A b_P - \eta a_R < 0,\ a_R b_A - \eta a_A < 0,\ b_A b_P > \eta^2$ | $E_c = \pm\frac{\pm 2}{3\sqrt{3}}\frac{\left(-a_R + \frac{\eta a_A}{b_A}\right)^{3/2}}{b_P - \frac{\eta^2}{b_A}}$<br>$P_3(E_c) = 0$ |
| Antipolar/antiferroelectric phase (AFE) | $A_s = \pm\sqrt{-\frac{a_A}{b_A}}$,<br>$P_s = 0$,<br>$W_s = 0$ | $f_A = -\frac{a_A^2}{4b_A}$<br>$f_A = min,\ a_A < 0$,<br>$a_R b_A - \eta a_A > 0$,<br>$\eta > \sqrt{b_A b_P}$ | $E_{c1} = \pm\left(a_R - \frac{a_A}{\eta}b_P\right)\sqrt{-\frac{a_A}{\eta}}$,<br>$P_{c1}(E_{c1}) = \pm\sqrt{-\frac{a_A}{\eta}}$<br>$E_{c2} = \frac{\pm 2}{3\sqrt{3}}\frac{\left(a_R - \frac{\eta a_A}{b_A}\right)^{3/2}}{\frac{\eta^2}{b_A} - b_P}$<br>$P_{c2}(E_{c2}) = \pm\sqrt{\frac{a_R b_A - \eta a_A}{3(\eta^2 - b_P b_A)}}$. |
| Polydomain ferroelectric state (PDFE) | $W_s = \pm\sqrt{-\frac{a_W}{b_W}}$,<br>$P_s = 0,\ A_s = 0$ | $f_W = -\frac{a_W^2}{4b_W}$<br>$f_W = min,\ a_W < 0$,<br>$a_R b_W - \eta_W a_W > 0$,<br>$\eta_W > \sqrt{b_W b_P}$ | $E_{c1} = \pm\left(a_R - \frac{a_W}{\eta_w}b_P\right)\sqrt{-\frac{a_W}{\eta}}$,<br>$P_{c1}(E_{c1}) = \pm\sqrt{-\frac{a_W}{\eta_W}}$<br>$E_{c2} = \frac{\pm 2}{3\sqrt{3}}\frac{\left(a_R - \frac{\eta a_A}{b_A}\right)^{3/2}}{\frac{\eta^2}{b_A} - b_P}$ |

| | | | $P_{c2}(E_{c2}) = \pm\sqrt{\frac{a_R b_A - \eta a_A}{3(\eta^2 - b_P b_A)}}$. |
|---|---|---|---|

## Appendix B. Hysteresis loops for the 5-nm Hf$_x$Zr$_{1-x}$O$_2$ nanoparticles

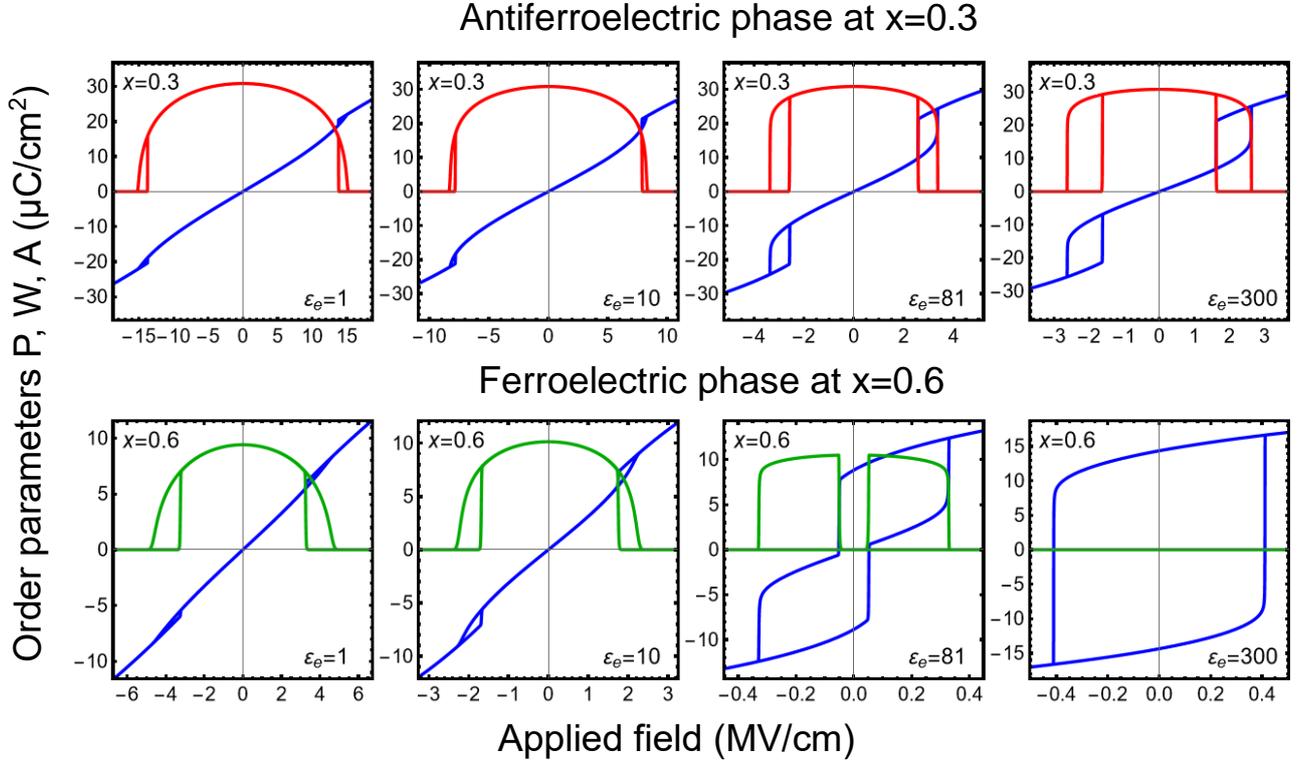

**FIGURE B1.** The quasi-static voltage dependences of the average polarization $P$ (the blue curves), antipolar order parameter $A$ (the red curves) and ferroelectric domains amplitude $W$ (the green curves), calculated for the Hf$_x$Zr$_{1-x}$O$_2$ nanoparticle, different values of $\varepsilon_e$ =1, 10, 81 and 300; and different $x$ =0.3 and 0.6 (the top and bottom rows, respectively). Corresponding values of $x$ and $\varepsilon_e$ are listed inside each plot. Other parameters: $R$ = 5 nm, $n_{1,2}$ =0.1 nm$^{-2}$ and $\Delta G = 0.1$ eV.

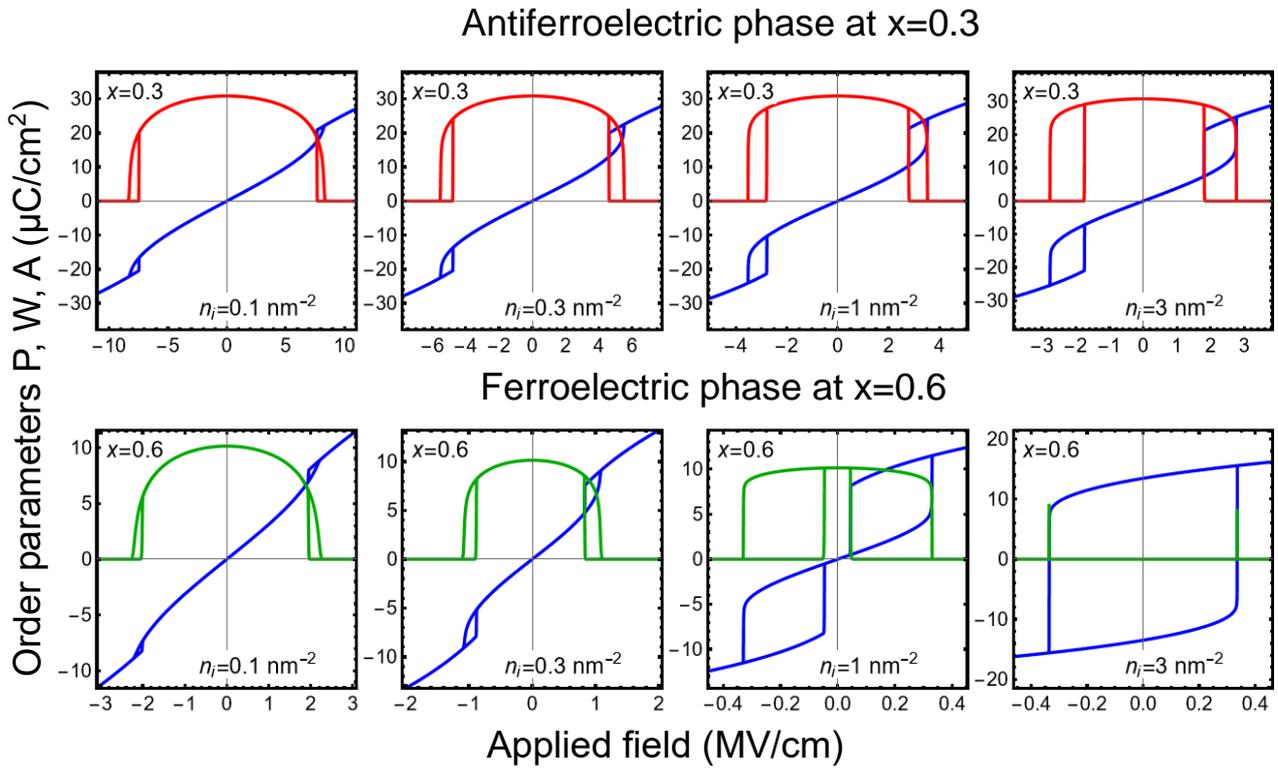

**FIGURE B2.** The quasi-static voltage dependences of the average polarization $P$ (the blue curves), antipolar order parameter $A$ (the red curves) and ferroelectric domains amplitude $W$ (the green curves), calculated for the Hf$_x$Zr$_{1-x}$O$_2$ nanoparticle, different values of $n_i$ varying from 0.3 nm$^{-2}$ to 3 nm$^{-2}$; and different $x$ =0.3 and 0.6 (the top and bottom rows, respectively). Corresponding values of $x$ and $\varepsilon_e$ are listed inside each plot. Other parameters: $R = 5$ nm, $\varepsilon_e = 10$ and $\Delta G = 0.1$ eV.